\documentclass[onecolumn,english,aps,floatfix,superscriptaddress,amsfonts,amssymb,superscriptaddress, pra]{revtex4}
\pdfoutput=1
\usepackage{graphicx}
\usepackage{amsmath}
\usepackage{latexsym}
\usepackage{epstopdf}
\usepackage{amsmath,amssymb}
\usepackage{amssymb,amsmath}
\usepackage{multirow}
\usepackage{mathtools}
\usepackage{color}
\newcommand{\beq}{\begin{equation}}
\newcommand{\eeq}{\end{equation}}

\newcommand{\bmat}[1]{\mathbf{#1}}

\def\xicor{|\boldsymbol{\xi}\rangle}
\def\xibarcor{\langle\boldsymbol{\bar{\xi}}|}
\def\bxi{\boldsymbol{\xi}}
\def\bxibar{\boldsymbol{\bar{\xi}}}
\def\bbeta{\boldsymbol{\eta}}
\def\bbetabar{\boldsymbol{\bar{\eta}}}
\begin{document}

\title{Return Amplitude after a Quantum Quench in the  XY Chain}

\author{Khadijeh Najafi  }
\affiliation{Department of Physics, Virginia Tech, Blacksburg, VA 24061, U.S.A}
\author{ M.~A.~Rajabpour}
\affiliation{  Instituto de F\'isica, Universidade Federal Fluminense, Av. Gal. Milton Tavares de Souza s/n, Gragoat\'a, 24210-346, Niter\'oi, RJ, Brazil}
\author{Jacopo Viti}
\affiliation{International Institute of Physics, UFRN,
Campos Universit\'ario, Lagoa Nova 59078-970 Natal, Brazil}
\affiliation{ECT, UFRN,
Campos Universit\'ario, Lagoa Nova 59078-970 Natal, Brazil}
\date{\today{}}

\begin{abstract}

 We determine an exact formula for the transition amplitude between any two arbitrary eigenstates of the local  $z$-magnetization operators in the quantum  XY  chain. We further use this formula to obtain an analytical expression for the return amplitude  of fully polarized states and the N\'eel state on a  ring of length $L$. Then, we investigate finite-size effects in the return amplitude: in particular quasi-particle interference halfway along the ring, a phenomenon that has been dubbed traversal~\cite{FE2016}. We show that the traversal time and the features of the return amplitude at the traversal time  depend on the initial state and on the parity of $L$. Finally, we briefly discuss non-analyticities in time of the decay rates in the thermodynamic limit $L\rightarrow\infty$, which are known as dynamical phase transitions.

\end{abstract}

\maketitle
\section{Introduction}
\label{sec:intro}
A prominent theoretical tool to investigate non-equilibrium phenomena in many-body systems is the  \textit{quantum quench}. In the original formulation of the problem~\cite{Silva}, a quantum system is prepared in the ground state of a translation invariant Hamiltonian depending on a control parameter. Such a parameter is suddenly modified (quenched) so that the unitary time evolution of the quantum state is governed by the post-quench Hamiltonian.

In low dimensions, tailored field theoretical~\cite{CC2006, CC2007, FM, SE, BES, Delfino2014, DV2017, KZ, TM},  free fermionic~\cite{Sac, Muk07, CEF2011, CEF2011a, CEF2011b, Igloi2012, Bucciantini2014b} and Bethe Ansatz techniques~\cite{QA, DeNa, DeNa2, FCEC14, Poz, Proz, MPC, BPC}  have been exploited for analytical investigation of the unitary dynamics after the quench. For reviews we refer to the special issue~\cite{CEM} and~\cite{Eis_rev, Pol_rev}.

There is a firm consensus~\cite{CEM} that macroscopic subsystems of a closed infinite system can equilibrate at late times to a stationary quantum density matrix. For one-dimensional integrable models, such a density matrix is a reduced density matrix obtained from a Generalized Gibbs Ensemble (GGE)~\cite{GGE_rigol, GGE_rigol2}. This recent theoretical study---especially after~\cite{KW}---has been motivated and boosted by the experimental implementation  with ultra-cold atoms of one-dimensional systems which are in almost perfect isolation from the external environment~\cite{coldatoms}.

In the framework pedagogically reviewed in~\cite{FE2016},  subsystem relaxation towards equilibrium can only occur if the volume of the system is infinite. In a finite system, the energy spectrum might be discrete and the wave function  recur~\cite{Bocchieri1957}, although with periods growing exponentially with the system size. More importantly, in an integrable model, finite size effects can prevent relaxation to a stationary state much earlier~\cite{Igloi2000} than a recurrence takes place. Suppose that an integrable spin chain is quantized on a ring of length $L$ and $v$ is a maximal~\cite{LR} effective propagation velocity for the quasi-particles.  Then, qualitatively, quasi-particles emitted in pairs~\cite{CC2006} can interfere halfway along the ring  at instants of time multiples of $T_{\text{tr}}=\frac{L}{2 v}$, which is $O(L)$. Following Sec. 5.2 in~\cite{FE2016}, we will call the time $T_{\text{tr}}$ around which  quasi-particle interference in finite volume prevent relaxation to a time-independent stationary state, the \textit{traversal time}.

An observable that diagnoses finite-size effects and  can provide a quantitative estimate  of $T_{\text{tr}}$ is the \textit{return amplitude} or fidelity~\cite{Quan2006, Yuan2007, Rossini2007, S2008, Zhong2011, Zanardi2011, MC12, Santos14, DeLuca14, PS14, Santos15, Mazza2016, Poz_echo, Poz_echo2}
\begin{equation*}
 F_{\Psi}(t)=|\langle\Psi|e^{-i\mathcal{H}t}|\Psi\rangle|,
\end{equation*}
where $|\Psi\rangle$ is the many-body initial state and $\mathcal{H}$, the post-quench Hamiltonian. 
In particular,  destructive or constructive quasi-particle interference is detectable in the logarithm of the return amplitude $\Gamma_{\Psi}(t)=-\frac{1}{L}\log F_{\Psi}(t)$; hereafter, we will refer to $\Gamma_{\Psi}$ as the decay rate. For the quenches considered in this paper, the decay rate relaxes in infinite volume at later times---or rather in a window $1\ll t\ll L$---to a constant~\cite{Z2010}, that is the limit 
 \begin{equation*}
 \bar{\Gamma}_{\Psi}\equiv\lim_{t\rightarrow\infty}\lim_{L\rightarrow\infty}-\frac{1}{L}\log F_{\Psi}
\end{equation*} 
 exists and can be  calculated. For finite large $L$ instead, when approaching  the traversal times, the amplitude of the decay rate oscillations  around $\bar{\Gamma}_{\Psi}$ increases; see  Fig.~\ref{fig:1} for an example that will be discussed in detail. At each traversal time, moreover, the decay rate exhibits  local maxima or minima, which signal that the system, being finite, cannot relax to a time-independent stationary state. Actually, after many traversals, the time average of the finite volume return probability $R_{\Psi}\equiv F_{\Psi}^2$  converges to its value in the diagonal ensemble~\cite{Z2010, F13, LV, Mazza2016}, $R_{\Psi}^{\text{DE}}$.
A correspondent decay rate in infinite volume can be extracted which  generally  differs~\cite{Z2010, F13} from $\bar{\Gamma}_{\Psi}$ defined earlier; cf.  Eq.~\eqref{decay_diag} in Sec.~\ref{sec:3} of this paper.

 Finite-size effects in the return amplitude have been investigated in 
 the last years~\cite{Happola, Montes2012, Sharma2012,  Rajak2014, KR2017, Jafari2017, Damski,Jafari2019}  seeking for a revival---i.e. an 
 instant at which $F_{\Psi}=O(1)$---of the initial wave function. However, as pointed out in~\cite{FE2016} and stressed once more in 
 the Sec.~\ref{sec:3} of this paper,  at the traversal time  there is usually not a \textit{bona fide} revival of the quantum state.  The return amplitude remains always exponentially  small with respect to the system size, i.e. $O(e^{-L})$ even for reasonably 
 large but finite systems.
 Indeed a necessary condition
 for a revival in a time which is $O(L)$ of  a state that has an exponential number of non-zero overlaps with the eigenenergy basis, 
 is the fact that the many-body spectrum, in a certain energy unit $\omega=O(1/L)$,  is integer spaced. 
This can happen, 
 for instance, in the continuum limit if non-interacting quasi-particles quantized on a cylinder  have a linear dispersion relation.
 In a lattice model however, the quasi-particles dispersion is  never linear on the whole Brillouin zone. The absence of  revivials 
 in the critical XY chain was investigated in detail  in~\cite{KR2017}. As a matter of fact, up to now, one can observe  physically relevant~\cite{B2017} 
 examples of revivals only within the context of conformal invariant field theories~\cite{DS2011, Cardy_echo}.

The decay rate of the return amplitude in the thermodynamic limit can also show non-analyticities in time that have 
been dubbed \textit{dynamical phase transitions}~\cite{Heyl2013} and are the focus of an intense 
research activity~\cite{Pollmann2010, Karrasch2013, AS, SB, Gullo2015, Lupo2016, Pollmann2016, Heyl2017, H1, H2, H3, S16, BD18, TH18, Sirker18, H4}; 
for a recent comprehensive survey we refer to the review~\cite{Heyl_rev}. The nature of such singularities and 
a possible link with the statistical mechanics are still unsettled~\cite{AS, Poz_echo}, although they have been 
observed experimentally~\cite{DPT_exp2,DPT_exp}; see also the viewpoint in~\cite{G17}.

In this paper, we will study both the finite size effects  and the non-analyticities in the thermodynamic 
limit for the return amplitudes in the XY chain~\cite{LMS}. We will focus on initial states $|\Psi\rangle$ that 
are eigenstates of the local $z$-magnetization operator; examples will include fully polarized states and 
the N\'eel state. These initial states are Gaussian---i.e. the Wick theorem holds~\cite{KRV2018a, BTC2018}---and allow
local relaxation in the thermodynamic limit to a stationary state~\cite{SC, BTC2018}; moreover, they are 
experimentally relevant~\cite{DPT_exp,DPT_Monroe}.

Our analytical results are based on a new formula for the matrix elements of the evolution operator associated 
with a quadratic fermionic Hamiltonian, which we will derive in Sec.~\ref{sec:1}; see Eq.~\eqref{Golden2}. In Sec.~\ref{sec:2}, we 
specialized this result to calculate return amplitudes in the XY chain. In Sec.~\ref{sec:3}, we provide detailed 
physical applications of the formalism and analyze the finite-size effects and the singularities in the thermodynamic 
limit in the decay rate of a fully polarized state and of the N\'eel state. We will then calculate the traversal times 
and show that they are initial state dependent; we will also determine the instants of time at which singularities in 
the thermodynamic limit of the decay rate show up. Our conclusions are gathered in Sec.~\ref{sec:4},  and an appendix completes the paper.

\section{Return amplitude in free fermionic chains}
\label{sec:1}
The aim of this section is the derivation of a formula to calculate the Return Amplitude (RA) for an arbitrary many-body 
state time-evolving  with a free fermionic Hamiltonian. We start by considering the following quadratic Hamiltonian 
in the real space ($T$ denotes transposition below)
\begin{eqnarray}\label{H1}\
\mathcal{H}_{\text{free}}=\textbf{c}^{\dagger}\textbf{A}\textbf{c}+\frac{1}{2}\textbf{c}^{\dagger}\textbf{B}\textbf{c}^{\dagger}+\frac{1}{2}\textbf{c}\textbf{B}^{T}\textbf{c}-\frac{1}{2}{\rm Tr}{\textbf{A}},
\end{eqnarray} 
where $\textbf{A}$ and $\textbf{B}$ are $L\times L$ real symmetric and antisymmetric matrices, 
respectively. The $L$-dimensional column vector $\textbf{c}$ contains the annihilation 
operators $c_j$ with $j=1,\dots,L$. Their Hermitian conjugates are $c_{i}^{\dagger}$ and canonical 
anticommutation relations hold. Transposition of column vectors is understood if required to define a 
fermionic quadratic form, such as in Eq.~\eqref{H1} or Eqs.~\eqref{eq_1}-\eqref{final_LE}. Following Balian 
and Brezin~\cite{Balian1969}, we introduce the $2L$-dimensional column vector $\boldsymbol{\gamma}\equiv(\boldsymbol{c}, \boldsymbol{c}^{\dagger})$. The 
evolution operator acts linearly on the vector $\boldsymbol{\gamma}$ as
\begin{equation}
 \label{lin_act}
 e^{i\mathcal{H}_{\text{free}} t}\boldsymbol{\gamma} e^{-i\mathcal{H}_{\text{free}} t}=\textbf{T}\boldsymbol{\gamma},
\end{equation}
where the $2L\times 2L$ matrix $\textbf{T}$ is obtained by applying the Baker-Campbell-Hausdorff formula (cf. appendix~A)
\begin{eqnarray}\label{mat_T}\
\textbf{T}=e^{-it \begin{pmatrix}
\textbf{A} & \textbf{B}\\
\textbf{-B} & \textbf{-A}\\
  \end{pmatrix}  }
\equiv\begin{pmatrix}
\textbf{T}_{11} & \textbf{T}_{12}\\
\textbf{T}_{21} & \textbf{T}_{22}\\
  \end{pmatrix}. 
\end{eqnarray} 
The $L\times L$ blocks $\textbf{T}_{\alpha\beta}$ ($\alpha,\beta=1,2$) can be written down explicitly if the free fermionic system is translation invariant; in the next section we will provide a physical example.
As anticipated in Sec.~\ref{sec:intro}, given a quantum state $|\Psi\rangle$, its RA is defined by 

\begin{equation}
 \label{fid}
 F_{\Psi}(t)=\left|\langle \Psi|e^{-i\mathcal{H}_{\text{free}}t}|\Psi\rangle\right|.
\end{equation}
If $|\Psi\rangle$ is an eigenstate of the local fermion  occupation numbers, a determinant representation for $F_{\Psi}$ can be obtained through the following route. First, we factorize the evolution operator $e^{-i\mathcal{H}_{\text{free}}t}$ \textit{\`a la} Balian-Brezin~\cite{Balian1969} as
\begin{eqnarray}\label{H2}\
e^{-i\mathcal{H}_{\text{free}}t}=e^{-\frac{1}{2}{\rm Tr}\textbf{Y}}e^{\frac{1}{2}\textbf{c}^{\dagger}\textbf{X}\textbf{c}^{\dagger}}e^{\textbf{c}^{\dagger}\textbf{Y}\textbf{c}}
e^{\frac{1}{2}\textbf{c}\textbf{Z}\textbf{c}},
\end{eqnarray} 
where $\textbf{X}$, $\textbf{Y}$, $\textbf{Z}$ can be calculated from the four blocks of the matrix $\textbf{T}$ in Eq.~\eqref{mat_T}
\begin{eqnarray}\label{X}\
\textbf{X}=\textbf{T}_{12}[\textbf{T}_{22}]^{-1},\hspace{0.5cm}\textbf{Z}=[\textbf{T}_{22}]^{-1}\textbf{T}_{21},\hspace{0.5cm}e^{\textbf{-Y}}=[\textbf{T}_{22}]^{T}.
\end{eqnarray} 
The Balian-Brezin factorization is briefly outlined in the ~appendix~\ref{app1}; here, we only remark that $\textbf{X}$ and $\textbf{Z}$ are complex antisymmetric matrices. The decomposition~\eqref{H2} is suitable for a coherent state representation of the RA,  for analogous manipulations of other quantum matrix elements see also~\cite{Mizusaki13}. Coherent fermionic states $|\bxi\rangle$ are introduced in the usual way through a set of $L$ anticommuting Grassmann numbers $\xi_i$, $i=1,\dots L$; namely
\begin{equation}
 |\boldsymbol{\xi}\rangle=e^{-\boldsymbol{\xi}\cdot\bmat{c^{\dagger}}}|\Omega\rangle,
\end{equation}
where $|\Omega\rangle$ is the vacuum defined by $\bmat{c}|\Omega\rangle=0$. Analogously, we can define the left coherent state as
 $\langle\boldsymbol{\bar{\xi}}|=\langle\Omega|e^{\boldsymbol{\bar{\xi}}\cdot\bmat{c}}$, where the set of Grassmann numbers $\boldsymbol{\bar{\xi}}$ is independent from the set $\boldsymbol{\xi}$.  Recalling  that Grassmann numbers anticommute with creation/annihilation operators, it is easy to verify that $\bmat{c}\xicor=\boldsymbol{\xi}\xicor$ and  analogously $\langle\boldsymbol{\bar{\xi}}|\bmat{c^{\dagger}}=\xibarcor\boldsymbol{\bar{\xi}}$. We finally also mention
 the resolution of the identity and the scalar product between two coherent states, i.e.
\begin{equation}
\label{coher_rel}
\int d\bxibar d\bxi e^{-\bxibar\cdot\bxi}\xicor\xibarcor=\boldsymbol{I},\quad\quad\xibarcor\bxi\rangle=e^{\bxibar\cdot\bxi},
\end{equation}
where $\textbf{I}$ is the $2^L\times 2^L$ identity matrix, acting on the fermionic Fock space and  $d\bxibar d\bxi\equiv d\bar{\xi}_n\dots d\bar{\xi}_1 d\xi_1\dots d\xi_L$. Let us now consider the amplitude
\begin{equation}
\label{amplitude}   
 f_{\Psi}(t)=\langle\Psi|e^{-it\mathcal{H}_{\text{free}}}|\Psi\rangle,
\end{equation}
where the state $|\Psi\rangle$ is an eigenstate of the local fermion occupation numbers, that is the overlaps $\langle\boldsymbol{\bar{\xi}}|\Psi\rangle=\prod_{j\in S_{\Psi}}\bar{\xi}_j$, where $S_{\Psi}\equiv\{i_1,\dots,i_m\}$ contains the lattice sites with fermion occupation one, for $1\leq i_1<i_2\dots<i_m\leq L$. Inserting two resolutions of the identity in the Eq.~\eqref{amplitude}, and using Eq.~\eqref{H2}, we obtain
\begin{equation}
\label{eq_1}
 f_{\Psi}(t)=e^{-\frac{1}{2}\text{Tr}(\textbf{Y})}\int d\bxibar d\bxi\int d\bbetabar d\bbeta~ e^{-\bxibar\cdot\bxi-\bbetabar\cdot\bbeta+\frac{1}{2}\bbetabar \textbf{X}\bbetabar+\frac{1}{2}\bxi \bf{Z}\bxi}
 \langle\Psi|\bbeta\rangle\langle\bbetabar|e^{\bmat{c^{\dagger}}\bf{Y}\bmat{c}}|\bxi\rangle\langle\bxibar|\Psi\rangle.
\end{equation}
The non-trivial matrix element $\langle\bbetabar|e^{\bmat{c^{\dagger}}\bf{Y}\bmat{c}}|\bxi\rangle$ in Eq.~\eqref{eq_1}, can be calculated by observing that, see also~\cite{Mizusaki13}, $e^{\bmat{c^{\dagger}}\bf{Y}\bmat{c}}|\bxi\rangle=|e^{\bf{Y}}\bxi\rangle$ and then using Eq.~\eqref{coher_rel}. Eventually, we end up with the Grassmann integral
\begin{equation}
\label{eq_2}
 f_{\Psi}(t)=e^{-\frac{1}{2}\text{Tr}(\bf{Y})}\int d\bxibar d\bxi\int d\bbetabar d\bbeta~ e^{-\bxibar\cdot\bxi-\bbetabar\cdot\bbeta+\frac{1}{2}\bbetabar \bf{X}\bbetabar+\frac{1}{2}\bxi \bf{Z}\bxi+\bbetabar e^{\bf{Y}}\bxi}\prod_{j\in S_{\psi}}\eta_{j}\prod_{j\in S_{\psi}}\bar{\xi}_j.
\end{equation}
Integrating Eq.~\eqref{eq_2} first over the Grassmann  variables $\bar{\xi}$ and  $\eta$ and then over $\xi$ and $\bar{\eta}$
 results in
\begin{align}
f_{\Psi}(t)&=(-1)^{(L+1)(L-m)}e^{-\frac{1}{2}\text{Tr}(\bf{Y})}\int d\xi_L\dots d \xi_1 d\bar{\eta}_L\dots d\bar{\eta}_1~e^{\frac{1}{2}\bbetabar \bf{X}\bbetabar+\frac{1}{2}\bxi \bf{Z}\bxi+\bbetabar e^{\bf{Y}}\bxi}\prod_{j\not\in S_{\psi}}\bar{\eta}_{j}\prod_{j\not\in S_{\psi}}\xi_{j}\nonumber\\
\label{final_LE}
&= e^{-\frac{1}{2} \text{Tr}(\textbf{Y})} \text{Pf}(\tilde{\textbf{M}}),
\end{align}
where the antisymmetric matrix $\tilde{\bf{M}}$ is obtained from
\begin{equation}
\label{matrixM}
 \textbf{M}=\begin{pmatrix}
    \textbf{X} & e^{\textbf{Y}}\\
    -e^{\textbf{Y}^T} & \textbf{Z}
   \end{pmatrix},
   \end{equation}
keeping only the $2m$ lines and columns $\{i_1,\dots,i_m\}$ and $\{i_1+L,\dots,i_m+L\}$. In practice,  these are the lines and columns in correspondence with the lattice sites occupied by a fermion.  We can finally  conclude (cf. Eq.~\eqref{fid}) 
\begin{equation}
\label{Golden2}
 F_{\Psi}(t)=\left|e^{-\frac{1}{2}\text{Tr}(\textbf{Y})}\text{Pf}(\tilde{\textbf{M}})\right|,
\end{equation}
which is an explicit expression for the RA and the main result of this section. The determinant representation in Eq.~\eqref{final_LE} can be also checked by observing that the initial state can be represented as
\begin{equation}
|\Psi\rangle=\int d\xi_{i_m}\dots d\xi_{i_1}|\bxi_{\Psi}\rangle,
\label{psi}
\end{equation}
where the coherent state $|\bxi_{\Psi}\rangle$ is obtained from the vector $\bxi_{\Psi}$ of components $\xi_{j}$ if $j\in S_{\Psi}$ and zero otherwise. The substitution of the expression of $|\Psi\rangle$ discussed above into Eq.~\eqref{amplitude}, leads to Eq.~\eqref{final_LE} .

 Finally, we emphasize  that the technique outlined in this section allows calculating the transition amplitude between two arbitrary fermionic states $|\Psi_i\rangle$ and $|\Psi_f\rangle$, once they are decomposed 
 into eigenstates of the local occupation number operators $c^{\dagger}_{j}c_j$. Indeed, let $|\Psi\rangle$ and $|\Psi'\rangle$  to be two of such eigenstates, 
 then, $S_{\Psi}=\{i_1,\dots,i_m\}$ and  $S_{\Psi'}=\{i_1',\dots,i_{k}'\}$ are the set of the lattice sites occupied by one fermion in $|\Psi\rangle$ and $|\Psi'\rangle$, respectively.
Consequently, to calculate $\langle\Psi'|e^{-i\mathcal{H}_{\text{free}}t}|\Psi\rangle$, we can use Eq.~\eqref{Golden2} where
the matrix $\tilde{\boldsymbol{M}}$ is obtained by keeping only the $m+k$ lines and columns $\{i_1', i_2',\dots, i_k'\}$ and $\{i_1+L,i_2+L,\dots,i_{m}+L\}$. 
An example will be discussed in Sec.~\ref{sec:3} (see Eq.~\eqref{gamma_flip}) for $m=L$ and $k'=0$.

\section{The XY chain: Balian-Brezin factorization}
\label{sec:2}
A systematic application of the Eq.~\eqref{Golden2} is given by the calculation of the return amplitude in the XY spin chain~\cite{LMS}.
The Hamiltonian of the model is
\begin{eqnarray}\label{HXY1}\
\mathcal{H}_{XY}=-\frac{J}{2}\sum_{j=1}^{L}\Big{[}\left(\frac{1+\gamma}{2}\right)\sigma_{j}^{x}\sigma_{j+1}^{x}+\left(\frac{1-\gamma}{2}\right)\sigma_{j}^{y}\sigma_{j+1}^{y}\Big{]}-\frac{h}{2}\sum_{j=1}^{L}\sigma_{j}^{z},
\end{eqnarray} 
where $\sigma_{j}^{\alpha}$ $(\alpha=x,y,z)$ are the Pauli matrices and $\gamma$ and $h$ are real parameters conventionally called anisotropy and magnetic field. In the following, we will assume $J>0$ (ferromagnetic model) and $h>0$, the latter restriction is not essential but it simplifies the discussion~\cite{DR14} when considering odd values of $L$.  Although Eq.~\eqref{Golden2} is valid also for open boundary conditions and can be used for numerical calculations in that case, from now on, we only focus on the periodic case: $\sigma_{j}^\alpha=\sigma_{j+L}^\alpha$.  

The XY chain reduces to the Ising spin chain and free fermions hopping on a lattice for $\gamma=1$ and $\gamma=0$, respectively.
Using the Jordan-Wigner transformation, the XY spin chain can be mapped into the quadratic fermionic Hamiltonian in Eq.~\eqref{H1}~\cite{LMS}. To this end,  one introduces the fermionic creation operators as
$c_j^{\dagger}=\prod_{l=1}^{j-1}\sigma_l^z\sigma_j^{+}$, and after nowadays standard manipulations Eq.~\eqref{HXY1} can be recast in the form
\begin{eqnarray}\label{XY-ff}
\mathcal{H}_{XY}=\frac{J}{2}\sum_{j=1}^{L} (c_j^{\dagger}c_{j+1}+\gamma c_j^{\dagger}c_{j+1}^{\dagger}+\text{H.c.})-\frac{h}{2}\sum_{j=1}^{L}(2c_j^{\dagger}c_j-1),
\end{eqnarray}
where we defined $c_{L+1}^{\dagger}\equiv-\mathcal{N}c_{1}^{\dagger}$ and $\mathcal{N}=\pm 1$ is the eigenvalue of the conserved parity operator
\begin{equation}
\label{parity}
P\equiv\prod_{j=1}^{L}\sigma_j^z.
\end{equation}
Consequently, the  Hamiltonian in Eq.~\eqref{XY-ff} can be rewritten as  Eq.~(\ref{H1}), if we consider
\begin{align}
\label{mata}
&[\textbf{A}]_{ij}=-\delta_{i,j} h+\frac{J}{2}(\delta_{i+1,j}+\delta_{i-1,j})-\frac{J\mathcal{N}}{2}(\delta_{i,1}\delta_{j,L}+\delta_{i,L}\delta_{j,1}),\\
\label{matb}
&[\textbf{B}]_{ij}=\frac{J\gamma}{2}(\delta_{i+1,j}-\delta_{i-1,j})+\frac{J\gamma\mathcal{N}}{2}(\delta_{i,1}\delta_{j,L}-\delta_{i,L}\delta_{j,1}).
\end{align}
It is worth noticing that up to a phase, the quantum state $|\Psi\rangle$, introduced in Sec.~\ref{sec:1} can be identified with a state where spins at positions $i_1,\dots,i_m$ have positive $z$-component. In the following, although this nomenclature is not widespread~\cite{CEF2011a}, we will call the even-parity ($\mathcal{N}=1$) subspace of the Hilbert space the Neveu-Schwarz (NS) sector and the odd-parity ($\mathcal{N}=-1$) subspace, the Ramond sector (R).
In order to apply the formalism introduced in Sec.~\ref{sec:1} we must spell out the Balian-Brezin factorization of the evolution operator in the XY chain. This is not difficult because the matrices $\boldsymbol{A}$ and $\boldsymbol{B}$  commute.
Indeed, they can be diagonalized by the unitary matrix $[\textbf{U}]_{mk}=\frac{1}{\sqrt{L}}e^{im\phi_k}$ with the quantization condition for the momenta
\begin{equation}
\label{quant_cont}
\mathcal{N}e^{iL\phi_k}=-1.
\end{equation}
The above formula can be easily derived after specializing the eigenvalue equation for $\mathbf{A}$ and $\mathbf{B}$ to the first component $i=1$ in Eqs.~\eqref{mata}-\eqref{matb}. Assuming from now on $J=1$, a simple calculation shows that the eigenvalues of these matrices denoted by $\lambda^{\textbf{A}}_k$ and $\lambda^{\textbf{B}}_k$ are
\begin{equation}
 \lambda_k^{\textbf{A}}=-h+\cos(\phi_k);\quad\lambda_k^{\textbf{B}}=i\gamma\sin(\phi_k);
\end{equation}
with $k=1,\dots,L$  and  $\phi_k=\frac{2\pi}{L}\left(k-\frac{\mathcal{N}+1}{4}\right)$. Then $2L\times 2L$ matrix equation,~\eqref{mat_T}, becomes a $2\times 2$ matrix equation for the eigenvalues, denoted by $\lambda_k^{\textbf{T}_{\alpha\beta}}$, of the four mutually commuting blocks $\boldsymbol{T}_{\alpha\beta}$ 
\begin{equation}
\label{mat_Exp}
\text{exp}\left[-it \begin{pmatrix}
\lambda^{\textbf{A}}_k & \lambda^{\textbf{B}}_k\\
-\lambda^{\textbf{B}}_k & -\lambda^{\textbf{A}}_k\\
  \end{pmatrix}\right]=\begin{pmatrix}
\lambda^{\textbf{T}_{11}}_k & \lambda^{\textbf{T}_{12}}_k \\
\lambda^{\textbf{T}_{21}}_k & \lambda^{\textbf{T}_{22}}_k \\
  \end{pmatrix}.
\end{equation}
The matrix exponential in Eq.~\eqref{mat_Exp} can be explicitly calculated. From the Eqs.~\eqref{X}, we obtain  the eigenvalues
of the matrices $\textbf{X}$, $\textbf{Z}$ and $\textbf{Q}\equiv e^{\textbf{Y}}$,
\begin{equation}
\label{eigen_BB}
 \lambda_k^{\textbf{X}}=-\frac{i\lambda_k^{\textbf{B}}}{i\lambda_k^{\textbf{A}}+\varepsilon_k\cot(t\varepsilon_k)},\quad\lambda_k^{\textbf{Z}}=-\lambda_k^{\textbf{X}},\quad\lambda_k^{\textbf{Q}}=\frac{1}{\cos(t\varepsilon_k)+i\frac{\lambda_k^{\textbf{A}}}{\varepsilon_k}\sin(t\varepsilon_k)},
\end{equation}
where $k=1,\dots, L$ and $\varepsilon_k$ is a function that coincides with the Bogoliubov quasi-particle  dispersion relation~\cite{Sach_book, Fran_book}, 
\begin{equation}
\label{dispersion-relation}
\epsilon_k=\sqrt{(\cos\phi_k-h)^2+\gamma^2\sin^2\phi_k}.
\end{equation}
Notice that the eigenvalues of the matrices $\textbf{X}$ and $\textbf{Z}$ and $\textbf{Q}$ in the Eq.~\eqref{eigen_BB} are not affected by the choice of the branch of the square root in Eq.~\eqref{dispersion-relation}, which we assume is positive. To apply Eq.~\eqref{Golden2}, which involves the matrix $\tilde{\textbf{M}}$, obtained by removing rows and columns from $\textbf{M}$, we also need an expression for both $\textbf{X}$ and $\textbf{Q}$. This is possible since they are both diagonalized by the discrete Fourier transform $\textbf{U}$ and therefore can be written as
\begin{equation}
\label{XQmat}
 [\textbf{X}]_{lj}=\frac{1}{L}\sum_{k=1}^L\lambda_k^{\textbf{X}}e^{ i\phi_k(l-j)};\quad [\textbf{Q}]_{lj}=\frac{1}{L}\sum_{k=1}^L\lambda_k^{\textbf{Q}}e^{ i\phi_k(l-j)}.
\end{equation}
It is useful to observe, cf. Eq.~\eqref{matrixM}, that the matrix $\textbf{Q}$ is symmetric, see also~\cite{KRV2018a}.

It should be emphasized that the formalism discussed here and in Sec.~\ref{sec:1}  does not require the determination of the fermionic representation of the spectrum of $\mathcal{H}_{XY}$ at finite $L$, which is actually not a straightforward task~\cite{XYfacchi,DR14}. Finally, we observe that unlike the case of the ground state~\cite{DR14}, the parity sector, to which the initial states considered in this paper belong, will be easily inferred from Eq.~\eqref{parity}. 

\section{Examples: finite-size effects and singularities in the thermodynamic limit}
\label{sec:3}
We present some analytic calculations of RA in the XY chain, based on Eq.~\eqref{Golden2}. We will focus on the initial states that are eigenstates of the local $\sigma_i^z$ operators. We consider initial states denoted by $|\boldsymbol{\mathcal{B}}\rangle$ that are built by repeating an elementary block $\mathcal{B}$ of contiguous spins along the chain. For instance, the state $|\boldsymbol{\uparrow}\rangle\equiv|\uparrow\uparrow\dots\uparrow\uparrow\rangle$ is a fully polarized state in the positive $z$-direction while the state $|\boldsymbol{\downarrow}\boldsymbol{\uparrow}\rangle\equiv|\downarrow\uparrow\dots\downarrow\uparrow\rangle$ is  the Ne\'el state.

It is also useful to introduce the decay rate
\begin{equation}
\label{gamma_def}
 \Gamma_{\boldsymbol{\mathcal{B}}}(t)=-\frac{1}{L}\log F_{\boldsymbol{\mathcal{B}}}(t),
\end{equation}
which, for our protocols, remains finite in the thermodynamic limit; see~\cite{Viti_in, JM_return} for important exceptions in the context of inhomogeneous initial states.
We note that a decay rate can also be defined for any quantum transition probability  between two states $|\Psi_i\rangle$ and $|\Psi_f\rangle$  on a chain of length $L$ as: $-\frac{1}{L}\log|\langle\Psi_f|\Psi_i(t)\rangle|$.
\subsection{Return Amplitude for the Fully Polarized Initial State}

The fully polarized initial state belongs to the NS sector of the Hilbert space of the XY chain, independently of the parity of $L$. It is therefore understood that $\phi_{k}=\frac{2\pi}{L}(k-1/2)$ and $k=1,\dots, L$ whenever needed. Applying Eq.~\eqref{Golden2} and observing that $|\text{Pf}(\textbf{M})|=1$, it follows
\begin{equation}
\label{fid_up}
 F_{\boldsymbol{\uparrow}}(t)=\left|e^{-\frac{1}{2}\text{Tr}(\textbf{Y})}\right|=\prod_{k=1}^L|\lambda_k^{\textbf{Q}}|^{-1/2},
\end{equation}
which implies
\begin{equation}
\label{decay_gamma}
 \Gamma_{\boldsymbol{\uparrow}}(t)=-\frac{1}{4L}\sum_{k=1}^L\log\left[1-\left(\frac{|\lambda_{k}^{\textbf{B}}|\sin(t\varepsilon_k)}{\varepsilon_k}\right)^2\right].
\end{equation}
In the limit $L\rightarrow\infty$, we can formally replace  $\sum_{k=1}^L\rightarrow\frac{L}{2\pi}\int_{0}^{2\pi} d\phi$ and obtain
\begin{equation}
\label{thermo_decay}
\lim_{L\rightarrow\infty}\Gamma_{\boldsymbol{\uparrow}}(t)=-\frac{1}{8\pi}\int_{0}^{2\pi}d\phi~\log\left[1-\left(\frac{|\lambda^{\textbf{B}}(\phi)|\sin(t\varepsilon(\phi))}{\varepsilon(\phi)}\right)^2\right].
\end{equation}
\subsubsection{Stationary Value of the Decay Rate}

As anticipated in Sec.~\ref{sec:intro}, the thermodynamic limit of the decay rate in the Eq.~\eqref{thermo_decay} approaches a constant for  large times
\begin{equation}
\label{gamma_mean}
\bar{\Gamma}_{\boldsymbol{\uparrow}}\equiv\lim_{t\rightarrow\infty}\lim_{L\rightarrow\infty}~\Gamma_{\boldsymbol{\uparrow}}(t).
\end{equation}
The value of the constant in the Eq.~\eqref{gamma_mean} can be obtained explicitly: for $h>1$ the logarithm in the Eq.\eqref{thermo_decay} can be expanded in a convergent power series for any value of $t$. The infinite time limit is then calculated by retaining only the time-independent terms in the power series of the logarithm or more formally applying the Riemann-Lebesgue lemma to drop all the oscillating contributions. This is of course equivalent to replace the large time limit with  the time average in Eq.~\eqref{gamma_mean}; it turns out after resuming all the time-independent terms
\begin{equation}
\label{mean_exact}
\bar{\Gamma}_{\boldsymbol{\uparrow}}=-\frac{1}{4\pi}\int_{0}^{2\pi}d\phi~\log\left[\frac{1}{2}\left(1+\sqrt{1-\frac{|\lambda^{\textbf{B}}(\phi)|^2}{\varepsilon^2(\phi)}}\right)\right].
\end{equation}
A similar result has been obtained in~\cite{S2008,Z2010} for a different decay rate in the Ising spin chain. Although derived for $h\geq 1$, 
the Eq.\eqref{mean_exact} holds for $h<1$ as well.  In the time window $1\ll t\ll L$, the decay rate $\Gamma_{\boldsymbol{\uparrow}}$ 
in Eq.~\eqref{decay_gamma} gets closer to Eq.~\eqref{mean_exact};  an example is given for $L=1000$, $\gamma=1/2$ and $h=1$ in 
Fig.~\ref{fig:1}: compare the blue oscillating curve with the dashed horizontal black line for times sufficiently smaller than the system size.

\subsubsection{Traversals}
\label{sec:tra_up}
\begin{figure}[t]
\includegraphics[width=0.5\textwidth]{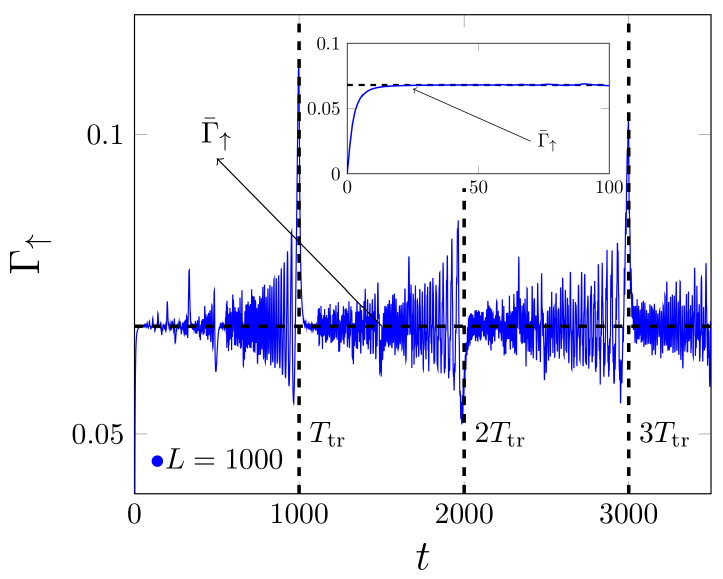}
\caption{The blue curve is the decay rate $\Gamma_{\boldsymbol{\uparrow}}(t)$ which is 
calculated from the Eq.~\eqref{decay_gamma} for $\gamma=1/2$ and $h=1$ at $L=1000$. For $1\ll t\ll L$  
the decay rate approaches the time average in the Eq.~\eqref{mean_exact} which is indicated by the dashed 
black horizontal line, see also the inset for a zoom in the relevant time window. Finite size 
effects are correctly predicted by Eq.~\eqref{sum_modes} for $v_{\text{in}}=\gamma$.  In agreement with 
the discussion in the main text for $t=T_{\text{tr}}$ the decay rate has a local maximum. Two other
traversal times are also shown in the figure. Notice that consistently at $t=2T_{\text{tr}}$ the decay rate has a local minimum.}
\label{fig:1}
\end{figure}

For large but finite $L$, the decay rate relaxes to its stationary value in the thermodynamic limit until a time $T_{\text{tr}}=O(L)$.
Within a qualitative quasi-particle picture~\cite{CC2006}, quasi-particles of opposite  momentum are emitted in pairs at $t=0$ and travel around the ring of length $L$ with an effective maximal velocity $v$. They can then interfere constructively or destructively at the time  $T_{\text{tr}}=\frac{L}{2v}$ when traversing half of the ring. According to the nomenclature introduced in Sec. 5.2 of Ref.~\cite{FE2016}, we will refer to $T_{\text{tr}}$ as a \textit{traversal time}.

A formal explanation of this effect and a quantitative estimation for $T_{\text{tr}}$ for  reasonably large $L$ is provided by the following argument.
If $L$ is large enough, we can approximate the quasi-particle dispersion $\varepsilon_k$, given in Eq.~\eqref{dispersion-relation} by a continuous function $\varepsilon(\phi)$. Now suppose that $\varepsilon(\phi)$ has an inflection point at $\phi=\phi_{\text{in}}\in[0,2\pi]$, i.e.
\begin{equation}
\label{lin_approx}
\varepsilon(\phi)=\varepsilon(\phi_{\text{in}})+v_{\text{in}}(\phi-\phi_{\text{in}})+O((\phi-\phi_{\text{in}})^3),
\end{equation} 
where we defined $v_{\text{in}}=\varepsilon'(\phi_{\text{in}})$.

For large $L$ , the interval $\Delta_{\phi}\equiv(\phi-\phi_{\text{in}})$ might still contain a macroscopic fraction of modes.  For all the discrete modes $\phi_k$ in the interval $\Delta_{\phi}$, we have $\phi_k-\phi_{\text{in}}=\frac{2\pi n_k}{L}+\delta_{0}$, where $0\leq \delta_0<\frac{2\pi}{L}$ and $n_k$ is an integer. Extracting the discrete modes in the interval $\Delta_{\phi}$ from the sum in the Eq.~\eqref{decay_gamma}, we can rewrite their contribution up to the first order in $\Delta_{\phi}$ as
\begin{equation}
\label{sum_modes}
-\frac{1}{4L}\sum_{\phi_k\in\Delta_{\phi}}\log\left[1-\left(\frac{|\lambda^{\textbf{B}}(\phi_{\text{in}})|\sin\bigl(t\varepsilon(\phi_{\text{in}})+\frac{2\pi n_k v_{\text{in}} t}{L}+\delta_{0}v_{\text{in}}t\bigr)}{\varepsilon(\phi_\text{in})}\right)^2\right].
\end{equation}
At times which are positive integer multiples of $T_{\text{tr}}=\frac{L}{2v_{\text{in}}}$, the elements of the sum in the Eq.~\eqref{sum_modes} are all equal, allowing larger fluctuation of the decay rate around the constant in Eq.~\eqref{mean_exact}. Rightly at the traversal time, the decay rate displays local maxima or minima which can be more pronounced depending on the validity of the linear approximation in the Eq.~\eqref{lin_approx} and the values of $\varepsilon(\phi_{\text{in}})$ and $\delta_0$ in the Eq.~\eqref{sum_modes}.

\begin{figure}[t]
\includegraphics[width=0.5\textwidth]{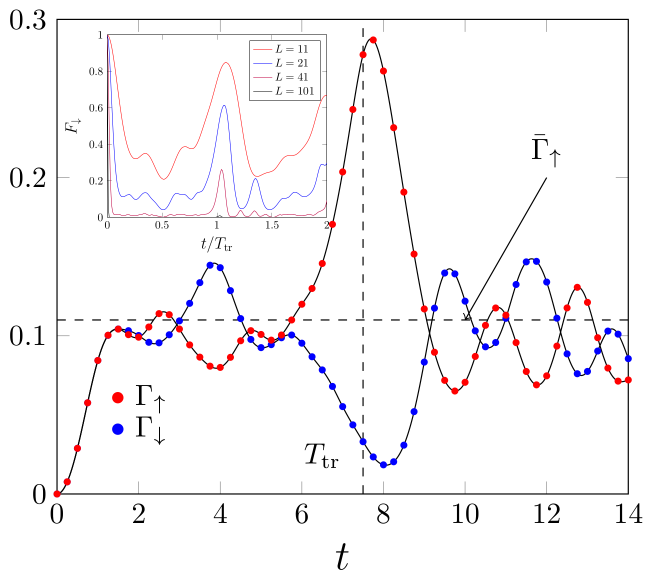}
\caption{The behaviour of the RA decay rate at the first traversal depends on the parity sector. The red and blue dots are results obtained for the decay rate of the two fully polarized states $|\boldsymbol{\uparrow}\rangle$ and $|\boldsymbol{\downarrow}\rangle$ by using the exact diagonalization at the critical Ising chain with $L=15$. The continuous lines are obtained from the Eq.~\eqref{gamma_def} with momenta in the NS (resp. R) sector when the state is fully polarized in the positive (resp. negative) $z$-direction. In agreement with the discussion in the main text $\Gamma_{\boldsymbol{\downarrow}}$ has a minimum at $t=T_{\text{tr}}$ ($v_{\text{in}}=1$) while $\Gamma_{\boldsymbol{\uparrow}}$ has a maximum.~\textit{Inset.} It shows $F_{\boldsymbol{\downarrow}}$ calculated from Eq.~\eqref{fid_up} with momenta in the R sector for different system sizes of the critical Ising chain ($\gamma=h=1$). At time $t=T_{\text{tr}}$ although the RA has a local maximum, there is no revival (according to the definition of~\cite{FE2016}) of the state $|\boldsymbol{\downarrow}\rangle$ .}
\label{fig:2}
\end{figure}

For instance, if $\varepsilon(\phi_{\text{in}})=0$ and $\delta_{0}=\frac{\pi}{L}$, as long as $\lim_{\phi\rightarrow\phi_{\text{in}}}|\lambda^{\textbf{B}}(\phi)|/\varepsilon(\phi)=1$, at time $t=T_{\text{tr}}$, each term in  the Eq.~\eqref{sum_modes} diverges logarithmically. In the actual spin chain, this logarithmic divergence is converted into a maximum of the decay rate,  which is clearly visible along the critical line  $h=1$  (here $\phi_{\text{in}}=0$,~$\varepsilon(0)=0$, $v_{\text{in}}=\varepsilon'(0)=\gamma$); examples are shown in the Fig.~\ref{fig:1} for $\gamma=1/2$  and for $\gamma=1$ in Fig.~\ref{fig:2}. Such a maximum of the decay rate indicates (see below the Eq.~\eqref{gamma_flip} for an explanation) that the time-evolved state is closer to a state fully polarized in the negative $z$-direction.

Eq.~\eqref{decay_gamma} for even $L$  and discrete momenta $\phi_k$ in the NS sector can also be applied to a fully negative polarized initial state, denoted hereafter by $|\downarrow\dots\downarrow\downarrow\rangle\equiv|\boldsymbol{\downarrow}\rangle$. However, if $L$ is odd, this is  no longer true. In such a case, the fully negative polarized initial state belongs to the R sector of the $XY$ chain (recall Eq.~\eqref{parity}) and the Eq.~\eqref{decay_gamma} gives its decay rate provided the discrete momenta $\phi_k$ are chosen accordingly. In particular, for odd $L$, along the critical line $h=1$, the decay rate $\Gamma_{\boldsymbol{\downarrow}}$ is minimum, contrary to $\Gamma_{\boldsymbol{\uparrow}}$,  when $t$ is close to $T_{\text{tr}}=\frac{L}{2\gamma}$. Indeed, for the R sector  $\delta_0=0$ and all the terms in the Eq.~\eqref{sum_modes} are now vanishing. We refer to Fig.~\ref{fig:2} where this prediction has been checked against exact diagonalization for the critical Ising spin chain with $L=15$  spins; see the blue curve.

Notice that, for odd $L$ and $h=1$, at the first traversal, $F_{\boldsymbol{\downarrow}}$ has a maximum; see  the inset in Fig.~\ref{fig:2}. Such a maximum, however, should not be confused with a revival of the quantum state $|\boldsymbol{\downarrow}\rangle$, which would happen~\cite{FE2016}, if the RA were of order one, independently of the system size. Instead, the Eq.~\eqref{sum_modes} suggests that for large enough $L$  at the traversal time, $F_{\boldsymbol{\downarrow}}(T_{\text{tr}})=O(e^{-L})$, unless only a finite  set of momenta do not fall in the interval $\Delta_{\phi}$; again see the inset in Fig.~\ref{fig:2}. The latter requirement is met, in physical models, only if the quasi-particle energy levels are all equally spaced, which is not the case for the $XY$ spin chain.

Finally, we remark that quantum revivals in the $XY$ chain of a fully polarized initial state are only possible in the limiting case $h=0$, $\gamma=1$. Indeed, for $h=0$ and $\gamma=1$,  the Hamiltonian in Eq.~\eqref{HXY1} trivializes and it is an exercise in the quantum mechanics to obtain
\begin{equation}
 F_{\uparrow}(t)=\left|\left(\cos(t/2)\right)^L+\left(i\sin(t/2)\right)^L\right|,
\end{equation}
which agrees with the Eq.~\eqref{gamma_def} for the same parameter values  and shows that the fully polarized state is reproduced at the revival times $T_{\text{rev}}=2\pi n$, $n=1,2,\dots$.

\subsubsection{Dynamical Phase Transitions}

At finite $L$, the RA of a quantum state vanishes when the time-evolved initial state is orthogonal to the final one (and vice-versa). For instance from the  Eq.~\eqref{fid_up}, it follows that  a necessary and sufficient condition for $F_{\boldsymbol{\uparrow}}=0$ is that it exists a $t_k$ such that $|\lambda_k^{\textbf{Q}}(t_k)|^{-1}=0$. However, for any value of $\gamma$,
\begin{equation}
 |\lambda_k^{\textbf{Q}}(t)|^{-1}=\sqrt{\cos(t\varepsilon_k)^2+\frac{(h-\cos(\phi_k))^2}{\varepsilon_k^2}\sin(t\varepsilon_k)^2},
\end{equation}
can vanish only for $h<1$ and  $\phi_{k_c}=\arccos(h)$  when
\begin{equation}
\label{critical_times}
t^{(n)}_{k_c}=\frac{\pi (n+1/2)}{\varepsilon(\arccos(h))},~\quad n=0,1,\dots
\end{equation}
The finite-size logarithmic divergences at $t^{(n)}_{k_c}$ are downgraded, for $L\rightarrow\infty$, to non-differentiable points of the decay rate in the Eq.~\eqref{thermo_decay}, 
which  have been termed dynamical phase transitions~\cite{Heyl2013}.  Indeed the same values for $t^{(n)}_{k_c}$ in the Eq.~\eqref{critical_times} 
were already obtained for the fully polarized initial state in the Ising spin chain ($\gamma=1$) in~\cite{Heyl2013}, 
relating them to so-called Fisher zeros~\cite{Fisher} of an analytically continued RA.  In summary,  Eq.~\eqref{critical_times} 
shows  that the generalization of the results in~\cite{Heyl2013} for a fully polarized initial state to $\gamma\not=1$ is straightforward and leads
to a qualitatively similar behaviour of the decay rate.

\subsection{Return Amplitude for the N\'eel Initial State}
From the formalism of Sec.~\ref{sec:1}, it is also possible to determine analytically the RA for the N\'eel state. We mention that the RA for a quench from the N\'eel state in the XY chain has been calculated in~\cite{AS, Mazza2016} at $\gamma=0$. It has been also worked out in the thermodynamic limit for the gapless XXZ spin chain in~\cite{Poz_echo, Poz_echo2} by a Quantum Transfer Matrix approach~\cite{K_rev, P14}; an earlier related numerical study was carried out, again for the XXZ spin chain,  in~\cite{AS}.

 According to the notations introduced at the beginning of this section, the N\'eel state is $|\boldsymbol{\downarrow\uparrow}\rangle$ and, assuming $L=2\ell$, it will belong to the NS sector if $\ell$ is even (i.e. $L=4n$) and to the R sector if $\ell$ is odd (i.e. $L=4n+2$). In order to apply Eq.~\eqref{Golden2}, we observe that the matrices $\tilde{\boldsymbol{X}}$,  $\tilde{\boldsymbol{Z}}=-\tilde{\boldsymbol{X}}$  and $\tilde{\boldsymbol{Q}}$ obtained from $\boldsymbol{X}$ and $\boldsymbol{Q}$ removing the odd lines and columns are again simultaneously diagonalized by the unitary matrix
$[\tilde{\boldsymbol{U}}]_{mk}=\frac{1}{\sqrt{\ell}}e^{2i\phi_k m}$, with $k,m=1,\dots,\ell$.  This can be proven as follows:

From Eq.~\eqref{XQmat}, we obtain the matrix
$[\tilde{\textbf{X}}]_{nm}=\frac{1}{2\ell}\sum_{k=1}^{2\ell}\lambda_k^{\textbf{X}}e^{2i\phi_k(n-m)}$. Let us apply such a matrix to the vector $\tilde{u}^{(k')}_m\equiv[\tilde{\boldsymbol{U}}]_{mk'}$, then, we obtain
\begin{equation}
\label{eigentilde}
 \sum_{m=1}^{\ell}[\tilde{\textbf{X}}]_{nm}\tilde{u}^{(k')}_m
 =\frac{1}{2\ell}\sum_{k=1}^{2l}\lambda^{\textbf{X}}_k\sum_{m=1}^{\ell} e^{2im(\phi_{k'}-\phi_k)}\frac{e^{2i\phi_{k}n}}{\sqrt{\ell}}=\frac{1}{2}(\lambda_{k'}^{\textbf{X}}+\lambda^{\textbf{X}}_{k'+\ell})\tilde{u}^{(k')}_n;
\end{equation}
where we used $u^{(k')}_n=u^{(k'+\ell)}_n$. An analogous calculation can be performed with the matrix $\tilde{\textbf{Q}}$ in Eq.~\eqref{XQmat},
 thus proving that $\tilde{\textbf{X}}$ and  $\tilde{\textbf{Q}}$ commute.

\begin{figure}[t]
\includegraphics[width=0.5\textwidth]{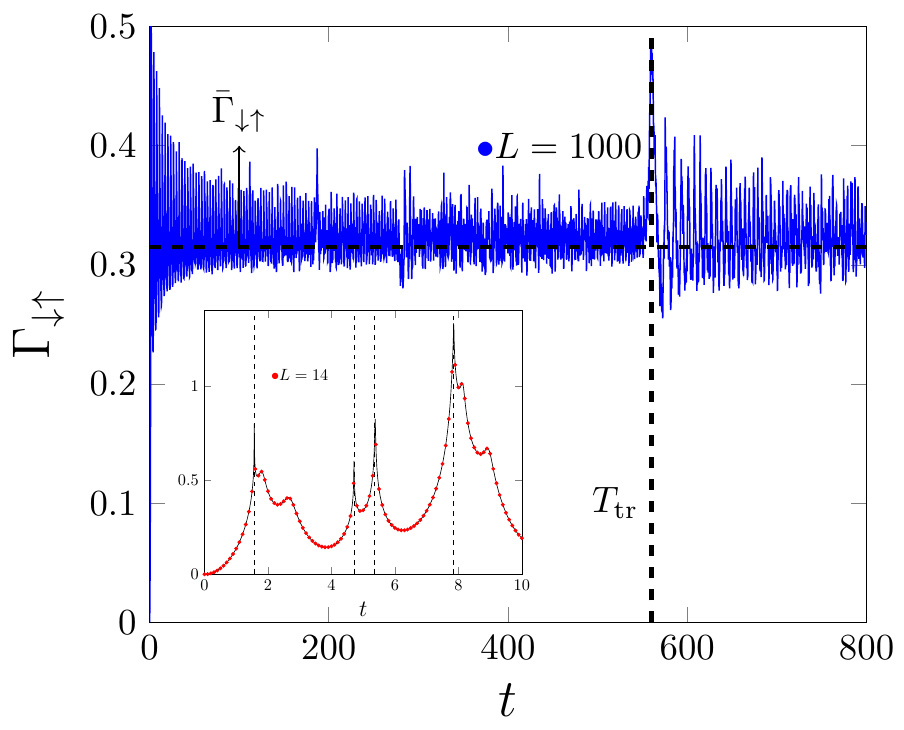}
\caption{The decay rate $\Gamma_{\boldsymbol{\downarrow\uparrow}}$ for $L=1000$ at $\gamma=1$ and $h=2$, see Eq.~\eqref{p2 Final}. The thick black dashed line is the constant obtained form the Eq.~\eqref{mean_exact_n} by calculating numerically the sum and the integral. The position of the first traversal has a very good agreement with the  Eq.~\eqref{sum_modes} and \eqref{vin}. As discussed in the main text, the traversal time is retarded and larger than $\frac{L}{2 v_{\text{max}}}$, where the $v_{\text{max}}$ the maximum quasi-particle velocity. ~\textit{Inset.} Red dots are obtained by calculating the N\'eel decay rate with exact diagonalization in the Ising chain with $h=2$ and $L=14$. The continuous black line is the Eq.~\eqref{p2 Final} with the momenta in the R sector. Peaks in the N\'eel decay rate correspond to the logarithmic divergences at the times given by Eq.~\eqref{dNeel1} and $t_{13}$ in the Eq.~\eqref{dNeel2}; see the main text.}
\label{fig:neel}
\end{figure}

Coming back to the  determination of the decay rate for the N\'eel state in the XY chain, we  recall that ${\rm Pf}(\tilde{\textbf{M}})=\pm\sqrt{\det{\tilde{\textbf{M}}}}$ and
\begin{equation}\label{detM10}
\det(\tilde{\textbf{M}})=\det(\tilde{\textbf{X}})\det(\tilde{\textbf{Z}}+\tilde{\textbf{Q}}\tilde{\textbf{X}}^{-1}\tilde{\textbf{Q}}),
\end{equation}
where the matrix $\tilde{\textbf{Z}}+\tilde{\textbf{Q}}\tilde{\textbf{X}}^{-1}\tilde{\textbf{Q}}$ in the Eq.~\eqref{detM10} is the Schur complement~\cite{Sc} of the block matrix $\tilde{\textbf{M}}$. Since all the matrices in the Eq.~\eqref{detM10} commute (cf. Eq.~\eqref{eigentilde} and the discussion above), we  obtain, from the Eq.~\eqref{Golden2}, the decay rate 
\begin{equation}\label{FT102}
\Gamma_{\boldsymbol{\downarrow\uparrow}}=\Gamma_{\boldsymbol{\uparrow}}-\frac{1}{2L}\sum_{k=1}^{\ell}\log\left|-(\lambda_{k}^{\tilde{\boldsymbol{X}}})^{2}+(\lambda_{k}^{\tilde{\boldsymbol{Q}}})^{2}\right|,
\end{equation}
where $\lambda_{k}^{\tilde{\textbf{X}}/\tilde{\textbf{Q}}}=\frac{1}{2}(\lambda_{k}^{\textbf{X}/\textbf{Q}}+\lambda^{\textbf{X}/\textbf{Q}}_{k+\ell})$ and $\Gamma_{\boldsymbol{\uparrow}}$ is given in the Eq.~\eqref{decay_gamma}. After some lengthy but straightforward  trigonometric manipulations, the Eq.~\eqref{FT102} can be eventually rewritten as 
\begin{equation}\label{p2 Final}
\Gamma_{\boldsymbol{\downarrow}\boldsymbol{\uparrow}}=-\frac{1}{2L}\sum_{k=1}^{\ell}\log \bigl[1-\cos^2(\bar{\phi}^b_{k})\sin^2(\epsilon_{k}^-t)-\sin^2(\bar{\phi}^b_{k})\sin^2(\epsilon_k^+t)\bigr],
\end{equation}
where we introduced the notations
\begin{equation}\label{bog}
\epsilon_{k}^{\pm}=\frac{\epsilon_{k}\pm \epsilon_{k+\ell}}{2},\quad
\bar{\phi}^b_{k}=\frac{\phi_{k}^b+\phi_{k+\ell}^b}{2},\quad\tan(\phi_{k}^b)=\frac{\gamma\sin(\phi_k)}{-h+\cos(\phi_{k})}. 
\end{equation}
When $h\leq 1$, for $\phi_k<\arccos(h)$ or $\phi_k>2\pi-\arccos(h)$ the angle $\phi^b_k$ in the Eq.~\eqref{bog} is defined as $\phi^{b}_k=\arctan\left(\frac{\gamma\sin(\phi_k)}{-h+\cos(\phi_k)}\right)\mp\pi$ respectively. Eq.~\eqref{p2 Final} with $L=4n$ and $h>1$  simplifies for $\gamma=0$ (i.e. free fermions hopping on the lattice), leading to~\cite{AS}
\begin{equation}
\label{neel_free}
F_{\boldsymbol{\downarrow}\boldsymbol{\uparrow}}^2=\prod_{k=1}^{\ell}\cos^2\left[ t\cos\left(\frac{2\pi(k-1/2)}{L}\right)\right],
\end{equation}
which is independent of $h$; see also~\cite{Mazza2016} for an analogous expression with open boundary conditions. We have checked the Eq.~\eqref{p2 Final} with exact diagonalization in the Ising spin chain up to $L=14$ spins; see the inset in Fig.~\ref{fig:neel} where the red dots are obtained by exact diagonalization and the continuous black line is Eq.~\eqref{p2 Final} with momenta in the R sector. In the thermodynamic limit  Eq.~\eqref{p2 Final} becomes 
\begin{equation}
\label{p2_th}
\lim_{L\rightarrow\infty}\Gamma_{\boldsymbol{\downarrow\uparrow}}=-\frac{1}{4\pi}\int_{0}^{\pi}d\phi\log\bigl[1-\cos^2(\bar{\phi}^b(\phi))\sin^2(\epsilon^-(\phi)t)
-\sin^2(\bar{\phi}^b(\phi))\sin^2(\epsilon^+(\phi)t)\bigr],
\end{equation}
where $\varepsilon^{\pm}(\phi)=\frac{1}{2}(\varepsilon(\phi)\pm\varepsilon(\phi+\pi))$ and $\bar{\phi}^{b}(\phi)=\frac{1}{2}(\phi^b(\phi)+\phi^b(\phi+\pi))$.
\subsubsection{Stationary Value of the Decay Rate}
 
At late times, the infinite volume N\'eel decay rate approaches a constant~\cite{Z2010} 
\begin{equation}
\label{gamma_n_mean}
\bar{\Gamma}_{\boldsymbol{\downarrow\uparrow}}\equiv\lim_{t\rightarrow\infty}\lim_{L\rightarrow\infty}~\Gamma_{\boldsymbol{\downarrow\uparrow}}(t).
\end{equation}
The value of such a constant can be calculated as briefly outlined above the Eq.~\eqref{mean_exact} for the fully polarized initial state. It turns out
\begin{equation}
\label{mean_exact_n}
\bar{\Gamma}_{\boldsymbol{\downarrow}\boldsymbol{\uparrow}}=\frac{1}{4\pi}\int_{0}^{\pi}d\phi \sum_{n=1}^{\infty}f_n(\bar{\phi}_b(\phi)),\quad f_{n}(x)=\frac{1}{2^{2n}n}\begin{pmatrix} 2n \\ n\end{pmatrix}\sin(x)^{2n}~{}_2F_{1}(1/2,-n,1/2-n,\text{cot}(x)^2),
\end{equation}
where ${}_2F_{1}$ is the Gauss hypergeometric function. Although convergence is rather slow, the sum and the integral in the Eq.~\eqref{mean_exact_n} can be evaluated numerically.
At $\gamma=0$, Eq.~\eqref{mean_exact_n} simplifies drastically and $\bar{\Gamma}_{\downarrow\uparrow}=\log(2)/2$. It is instructive to compare this result with the one that could be obtained taking the time-average of the finite volume return probability in Eq.~\eqref{neel_free}. Such a 
time average corresponds to the value of the return probability in the diagonal ensemble~\cite{Mazza2016}
\begin{equation}
\label{return_DE}
 R_{\downarrow\uparrow}^{\text{DE}}\equiv\lim_{T\rightarrow\infty}\frac{1}{T}\int_{0}^{T}dt~F_{\boldsymbol{\downarrow}\boldsymbol{\uparrow}}^2=\frac{1}{2^{L/2}}.
\end{equation}
Then we can extract the infinite volume decay rate
\begin{equation}
\label{decay_diag}
\lim_{L\rightarrow\infty}-\frac{1}{2L}\log(R_{\downarrow\uparrow}^{\text{DE}})=\log(2)/4
\end{equation}
 which is half of Eq.~\eqref{mean_exact_n} at $\gamma=0$. As anticipated in the Introduction, this example shows clearly that the thermodynamic limit and the infinite time limit do not commute when calculating the return probability and its  decay rate.

 \subsubsection{Traversals}
 Qualitatively~\cite{FE2016},  at the first  traversal time $T_{\text{tr}}=O(L)$ the decay rate in Eq.~\eqref{p2 Final}  stops relaxing towards the constant in Eq.~\eqref{mean_exact_n}. We can estimate $T_{\text{tr}}$ as follows: According to the discussion around Eq.~\eqref{lin_approx}, we look for inflection points of  the functions $\varepsilon^{\pm}$ in Eq.~\eqref{p2_th}. It turns out that $\varepsilon^{-}$ has, irrespectively of $\gamma$ and $h$, an inflection point at $\phi_{\text{in}}=\pi/2$ where $\bar{\phi}^{b}=0$. Therefore, quite interestingly, Eq.~\eqref{sum_modes} can be applied  as if on the critical line $h=1$ for the fully polarized initial state, upon replacing  $\varepsilon\rightarrow\varepsilon^{-}$ and
\begin{equation}
\label{vin}
v_{\rm{in}}=\left.\frac{d\varepsilon^-(\phi)}{d\phi}\right|_{\phi=\frac{\pi}{2}}=\frac{h}{\sqrt{h^2+\gamma^2}}.
\end{equation}
 Then for $L=4n$ and momenta $\phi_k$ in the NS sector $\delta_0=\frac{\pi}{L}$, the Eqs.~\eqref{sum_modes} and~\eqref{vin} predict 
 a maximum of the decay rate for $\Gamma_{\boldsymbol{\downarrow\uparrow}}$ at the traversal time $T_{\text{tr}}=\frac{L}{2 v_{\text{in}}}$. The 
 same conclusion applies for $L=4n+2$ and momenta $\phi_k$ in the R sector. These predictions have been checked in the Fig.~\ref{fig:neel} for $\gamma=1$ and $h=2$. In particular, we note that the location of the first traversal time for the N\'eel state is retarded with respect to the one of the fully polarized state. We can then conclude that $T_{\text{tr}}$ is state-dependent, and bounded by $\frac{L}{2v_{\text{max}}}$ with $v_{\text{max}}$  the maximum allowed quasi-particle dispersion given in Eq.~\eqref{dispersion-relation}. 

\subsubsection{Dynamical Phase Transitions}
\begin{figure}[t]
\includegraphics[width=0.5\textwidth]{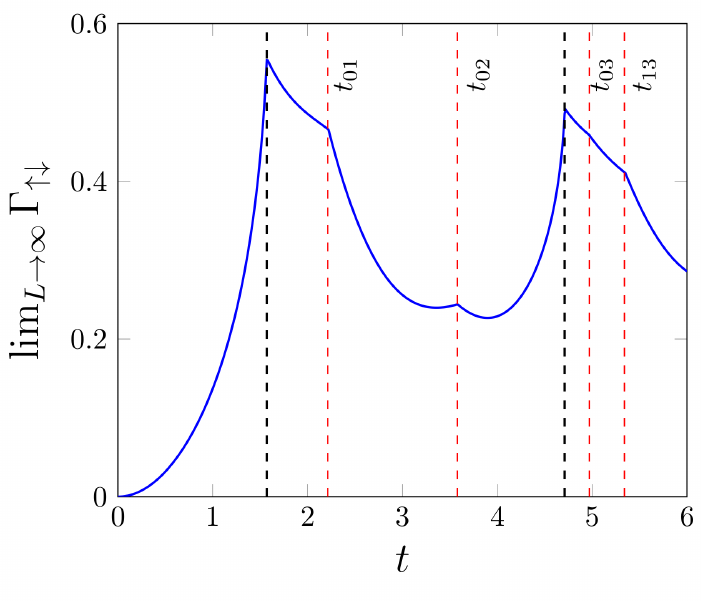}
\caption{The thermodynamic limit  $\lim_{L\rightarrow\infty}\Gamma_{\boldsymbol{\downarrow\uparrow}}(t)$ given in the Eq.~\eqref{p2_th} for the Ising spin chain ($\gamma=1$) with $h=2$. Non-differentiable points are the so-called dynamical phase transitions. Dashed black lines are obtained from the first two terms ($n=0,1$) of the series in Eq.~\eqref{dNeel1} while dashed red lines indicate the first four terms in the series of the Eq.~\eqref{dNeel2}. }
\label{fig:3}
\end{figure}
We can determine the instants of time at which the finite-size decay rate $\Gamma_{\boldsymbol{\downarrow\uparrow}}(t)$ is logarithmically divergent. We observe that the argument of the logarithm in the Eq.~\eqref{p2 Final} can be rewritten as
\begin{equation}
\label{arg_log}
\cos^2(\bar{\phi}^b_{k})\cos^2(\epsilon_{k}^-t)+\sin^2(\bar{\phi}^b_{k})\cos^2(\epsilon_k^+t)
\end{equation}
which can vanish if and only if each of the two squares in the sum vanishes. In particular,
\begin{itemize}
\item If $h>1$, $-\pi/2<\bar{\phi}^b_k<\pi/2$, then we can  have $\bar{\phi}_{k_c}^b=0$  for $\phi_{k_c}=\{0,~\pi/2,~\pi\}$ in the interval $[0,\pi]$. Since $\varepsilon^-(\phi)$ is zero for $\phi=\pi/2$ and $\varepsilon^-(\pi)=-\varepsilon^-(0)=1$ for $h>1$, we obtain a first series of singularities at
\begin{equation}
\label{dNeel1}
t^{(n)}_{k_c}=\frac{\pi(n+1/2)}{\varepsilon^-(\phi_{k_c})}=\pi (n+1/2),\quad n=0,1,\dots
\end{equation}  
with $\phi_{k_c}=\pi$. However, this is not enough, there might exist values of the angle  $\phi_{k^*}\in[0,\pi]$ such that
\begin{equation}
\label{dNeel2}
\varepsilon^-(\phi_{k^*})t_{nm}=(n+1/2)\pi,
~~\varepsilon^+(\phi_{k^*})t_{nm}=(m+1/2)\pi\Rightarrow
\frac{\varepsilon^-(\phi_{k^*})}{\varepsilon^+(\phi_{k^*})}=\frac{2n+1}{2m+1},~~n,m\in\mathbb Z
\end{equation}
and thus leading to a dynamical phase transition at $t=t_{nm}>0$. This is illustrated for the Ising spin chain with $h=2$ in the Fig.~\ref{fig:3}: black dashed lines denote  the first two terms in the series \eqref{dNeel1} $(n=0,1)$ while  red dashed lines indicate the first four singularities predicted by Eq.~\eqref{dNeel2}. 
\item If $h\leq 1$, we can have $-\pi/2\leq\bar{\phi}^b_k\leq\pi/2$. However  $\bar{\phi}^b_{k_c}=0$ only for $\phi_{k_c}=\pi/2$ while $\bar{\phi}^b_{k_c}=\mp \pi/2$ for $\phi_{k_c}=0,\pi$ respectively. Moreover, since for $h\leq 1$, $\varepsilon^{-}(\pi/2)=0$ and $\varepsilon^{+}(0)=\varepsilon^{+}(\pi)=1$, we obtain a series of dynamical phase transitions at times
\begin{equation}
t^{(n)}_{k_c}=\frac{\pi (n+1/2)}{\varepsilon^+(\phi_{k_c})}=\pi (n+1/2),~n=0,1\dots
\end{equation}
which are the same as the ones given in Eq.~\eqref{dNeel1} (again $\phi_{k_c}=\pi$). If $h\leq 1$, then, the argument leading to Eq.~\eqref{dNeel2} also remains intact.
\end{itemize}
This analysis supports the original claim~\cite{AS}  that dynamical phase transitions in the N\'eel decay rate  are not specific of quenches crossing the critical line $h=1$; they rather occur both quenching into the paramagnetic ($h>1$) or ferromagnetic phase ($h\leq 1$). Moreover, the location of the singularities depends non-trivially on the quasi-particle spectrum, see the Eq.~\eqref{dNeel2}. 

Finally, we observe that for finite $L$, the critical momenta $\phi_{k_c}$ and $\phi_{k^*}$ in the Eqs.~\eqref{dNeel1} and ~\eqref{dNeel2} might not be allowed by the NS or R quantization. In such a case the finite size Neel decay rate $\Gamma_{\downarrow\uparrow}$ will not be singular at the corresponding critical times. This is again shown in the inset of the Fig.~\ref{fig:neel} when $\gamma=1$ and $h=2$. 
For $L=14$, in the Ramond sector one can verify that $\phi_{k_c}=\pi$ is an allowed angle and therefore the N\'eel decay rate will be logarithmically divergent at times given in Eq.~\eqref{dNeel1}. In the inset of the Fig.~\eqref{fig:neel}, there is also a visible peak of the the decay rate at $t_{13}=5.349\dots$
given in Eq.~\eqref{dNeel2}, see also the Fig.~\ref{fig:3}. It is indeed possible to show that $\phi_{k=5}$  
is close (less than half of a percent) to the angle $\phi_{k^*}$, i.e. solution of the Eq.~\eqref{dNeel2}, for $n=1$ and $m=3$.

\subsection{The Flip Amplitude}

The method introduced in Sec.~\ref{sec:1}, allows also determining the decay rate of the transition probability between the two fully polarized initial states: $|\boldsymbol{\uparrow}\rangle$  and $|\boldsymbol{\downarrow}\rangle$.  The transition probability is zero for odd $L$ since the two states belong to different sectors of the Hilbert space, whereas for $L$ even and $\phi_k$ in the NS sector (observe that $\lambda_k^{\mathbf{Z}}=\frac{\varepsilon_k\lambda_k^{\textbf{X}}}{i\lambda_k^{\textbf{B}}\sin(t\varepsilon_k)}$)
\begin{equation}
\label{gamma_flip}
\Gamma_{\text{flip}}(t)\equiv-\frac{1}{L}\log\Bigl(|\langle \boldsymbol{\downarrow}|e^{-i\mathcal{H}_{XY}t}|\boldsymbol{\uparrow}\rangle|\Bigr)=-\frac{1}{2L}\sum_{k=1}^L\log\left|\frac{\lambda_{k}^{\textbf{B}}\sin(t\varepsilon_k)}{\varepsilon_k}\right|.
\end{equation}

 The analysis of the traversals parallels the one for the fully polarized initial state given 
 in Sec.~\ref{sec:tra_up}, therefore we will not repeat it here. We limit to observe that along
 the critical line $h=1$, at the first traversal time $T_{\text{tr}}=\frac{L}{2\gamma}$ 
 the decay rate $\Gamma_{\text{flip}}(T_{\text{tr}})$ has a minimum. This signals that the time evolved state
$e^{-i\mathcal{H}_{XY}t}|\boldsymbol{\uparrow}\rangle$  is closer, although still exponentially far for large $L$, to a state fully polarized in the negative $z$-direction. 

\subsubsection{Dynamical Phase Transitions}
\begin{figure}[t]
\includegraphics[width=0.5\textwidth]{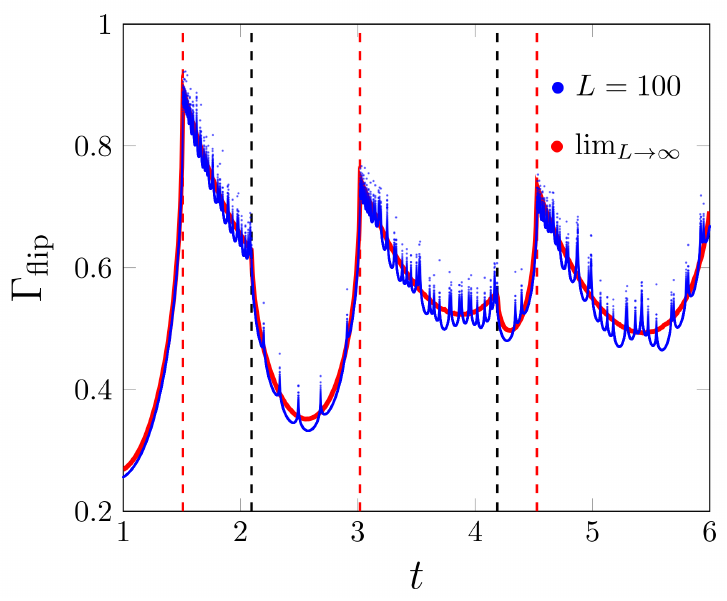}
\caption{Decay rate e.g. $\Gamma_{\text{flip}}(t)$ for $h=1/2$, $\gamma=2$, see the Eq.~\eqref{gamma_flip}. The blue curve is obtained by summing $L=100$ terms in the Eq.~\eqref{gamma_flip}. At $t=t^{(n)}_k$ the logarithmic singularities in the amplitude are visible~(see Eq.~\eqref{series_flip}). The red curve is obtained from the thermodynamic limit $L\rightarrow\infty$ of Eq.~\eqref{gamma_flip}, where we did the replacement $\sum_{k=1}^L\rightarrow\frac{L}{2\pi}\int_{0}^{2\pi}$---strictly speaking the sum is not convergent, as a Riemann integral, if $t\geq t_{\min}$. In the thermodynamic limit, there are still singularities at multiples of $t_{\min}\equiv\frac{\pi}{\max_{\phi\in[0,2\pi]}\varepsilon(\phi)}$, dashed red lines, and  $t_{\text{edge}}\equiv\frac{\pi}{\varepsilon(\pi)}$, dashed black lines.}
\label{fig:4}
\end{figure}
 We finally discuss dynamical phase transitions for the flip amplitude; to our best knowledge, this case has been never studied analytically before.
 At finite $L$, the decay rate $\Gamma_{\text{flip}}$ in the Eq.~\eqref{gamma_flip}  is  logarithmically divergent at
\begin{equation}
\label{series_flip}
t^{(n)}_k=\frac{n\pi}{\varepsilon_k},~\quad n=1,2,\dots
\end{equation}
and $k=1,2,\dots L$; see the blue curve in the Fig.~\ref{fig:4}, obtained for $\gamma=2$, $h=1/2$ and $L=100$. A similar behaviour has been also found in~\cite{AS} for a different physical setting:  a quench from the N\'eel state in a free fermionic chain with staggered magnetic field; cf. Fig. 4 there. 

In the thermodynamic limit $L\rightarrow\infty$, we can formally replace $\sum_{k=1}^L\rightarrow\frac{L}{2\pi}\int_{0}^{2\pi}$ in the Eq.~\eqref{gamma_flip}; the result is shown by the red curve in the Fig.~\ref{fig:4}.  We then conjecture that  non-analyticities in the integral $\lim_{L\rightarrow\infty}\Gamma_{\text{flip}}(t)$ occur at discrete times which are multiples of $t_{\min}\equiv\frac{\pi}{\max_{\phi\in[0,2\pi]}\varepsilon(\phi)}$, which is the first singularity in the series of Eq.~\eqref{series_flip}, and of $t_{\text{edge}}\equiv\frac{\pi}{\varepsilon(\pi)}$. The latter are 
in correspondence with a single-particle energy at the edge of the spectrum. We have checked the last statement numerically in
the XY chain, however we do not have  a satisfactory analytical understanding of such a claim. An example is again given 
in the Fig.~\ref{fig:4} for $\gamma=2$ and $h=1/2$: dashed red lines denote multiples of $t=nt_{\min}$ ($n=1,2,3$), 
while dashed black lines correspond to $t=nt_{\text{edge}}$ ($n=1,2$). Finally, notice that if the initial state is a 
cat state $|\text{cat}\rangle=2^{-1/2}\left(|\boldsymbol{\uparrow}\rangle+|\boldsymbol{\downarrow}\rangle\right)$, then
 \begin{equation}
 \label{exp}
\lim_{\ell\rightarrow\infty}-\frac{1}{2\ell}\log\left(|\langle\text{cat}| e^{-i\mathcal{H}_{XY}t}|\text{cat}\rangle|\right)=\min\{\Gamma_{\boldsymbol{\uparrow}}(t),\Gamma_{\text{flip}}(t)\}.
 \end{equation}
where $L=2\ell$. One can find the Eq.~\eqref{exp} simply by observing that the interference term is bounded for large $L$ by $2e^{-L(\Gamma_{\boldsymbol{\uparrow}}+\Gamma_{\text{flip}})}$, which is always exponentially small in the system length compared to $e^{-2L\min\{\Gamma_{\boldsymbol{\uparrow}},\Gamma_{\text{flip}}\}}$. This  non-equilibrium protocol was considered in the experiment~\cite{DPT_exp} and proposed theoretically in~\cite{S16} for an Ising spin chain with the long-range couplings. However, no fully analytical prediction was presented so far.

\section{Conclusions}
\label{sec:4}
In this paper, we studied the return amplitude in the XY spin chain for the eigenstates of the local $z$-component magnetization operator. We obtained analytic expressions for the fully polarized states and the N\'eel state, which generalize the  previous results in~\cite{Quan2006, AS, Mazza2016} to arbitrary values of the anisotropy parameter $\gamma$. These formulas have been derived from a  determinant representation for the matrix elements of the evolution operator of a quadratic fermionic Hamiltonian, see Eq.~\eqref{Golden2}.  We then focussed on the analysis of the  finite-size effects in the return amplitude showing that they are signalled by the so-called traversals~\cite{FE2016}, whose features depend also on the parity of the length of the spin chain. In particular, at the first traversal, the decay rate might show a maximum or a minimum depending on the quantization sector (NS or R) to which the initial state belongs. Analogously, we provided evidence that traversal times are also initial state dependent. Our
results have been tested with exact diagonalization methods for the Ising spin chain up to $L=15$. We hopefully made it clear that at a traversal time the return amplitude is expected to be $O(e^{-L})$ and therefore such a finite-size effect does not lead to a revival of the quantum state; in agreement with the discussion in~\cite{FE2016}. Finite-size effects in the evolution of the entanglement entropy~\cite{Fagotti2011, KRV2018a, BTC2018} could be in principle studied similarly. Our approach is also suitable to extract overlaps between the initial states examined in this paper and the eigenstates of the XY Hamiltonian. Analytical results for the overlaps obtained in~\cite{Mazza2016} for the N\'eel state could be extended to $\gamma\not=0$ as well.

We have moreover analyzed the thermodynamic limit of the logarithm of the return amplitude, namely the decay rate, and identified analytically the instants of time at which it might develop non-analyticities. For the N\'eel decay rate, singularities in time do not follow a periodic pattern and they appear independently of the final Hamiltonian. Our results complement in this respect the numerical study of~\cite{AS} and the Bethe Ansatz analysis in~\cite{Poz_echo, Poz_echo2}.

The analytical results for $f_{\Psi}(t)=\langle \Psi| e^{-i\mathcal{H}_{XY}t}|\Psi\rangle$ contained in this paper, 
could be also exploited to extract, upon analytic continuation to imaginary time $t=i\tau$,  universal boundary entropies 
along the critical line $h=1$~\cite{AL}. We hope to come back on this problem in the near future.
\newline

\textbf{Acknowledgements.}
KN  acknowledges the supports by National
Science Foundation under Grant No. PHY-1620555 and DOE grant DE-SC0018326. MAR acknowledges the support from CNPq. JV thanks Rodrigo Pereira for a discussion.

\appendix
\section{Balian Brezin factorization}
\label{app1}
In this Appendix we review the Balian Brezin factorization~\cite{Balian1969}. Consider a  \textit{complex} quadratic fermionic form  (again transposition of column vectors is understood)
\begin{equation}
\label{qf}
 \mathcal{H}=\frac{1}{2}\boldsymbol{\gamma}\textbf{H}\boldsymbol{\gamma},
\end{equation}
where the  $2L$-dimensional column vector $\boldsymbol{\gamma}=(\bmat{c}, \bmat{c}^{\dagger})$ and the operators $c^{\dagger}_i$ and $c_j$  obey canonical anticommutation relations. The components of the vector  $\boldsymbol{\gamma}$  satisfy
\begin{equation}
 \{\gamma_i,\gamma_j\}=\sigma_{ij},~\quad\sigma=\begin{pmatrix}0 & \bmat{1}\\ \bmat{1} & 0\end{pmatrix}.
\end{equation}
It is also useful to observe that $\textbf{H}$ which in \eqref{qf} is a complex matrix can be taken complex antisymmetric. If $\textbf{H}$ is not antisymmetric we can write $\textbf{H}=\textbf{H}_a+\textbf{H}_s$, being $\textbf{H}_{a/s}$ the antisymmetric/symmetric parts of it. Now substituting into Eq.~\eqref{qf} we get
\begin{equation}
 \boldsymbol{\gamma} \textbf{H}\boldsymbol{\gamma}=\boldsymbol{\gamma} \textbf{H}_a\boldsymbol{\gamma}+\frac{1}{2}\text{Tr}[\textbf{H}\sigma],
\end{equation}
and all the formulas that follow have to be modified accordingly.
The transformation $\mathcal{F}=e^{\mathcal{H}}$ acts linearly on the fermion $\boldsymbol{\gamma}$. Indeed, we  can apply Baker-Campbell-Hausdorff formula and the commutation relations to prove that
\begin{equation}
\label{int_rel}
\mathcal{F}^{-1}\boldsymbol{\gamma}\mathcal{F}=\textbf{T}\boldsymbol{\gamma},
\end{equation}
being $\textbf{T}=e^{\sigma \textbf{H}}$. The matrix $\textbf{T}$ satisfies $\textbf{T}\boldsymbol{\sigma}\textbf{T}^{T}=\boldsymbol{\sigma}$.

 Let us now consider two transformations $\mathcal{F}_1$ and $\mathcal{F}_2$ of the same type as in the Eq.~\eqref{qf}. Their composition applied to $\gamma$ yields
\begin{equation}
 (\mathcal{F}_1\mathcal{F}_2)^{-1}\boldsymbol{\gamma}(\mathcal{F}_1\mathcal{F}_2)=\textbf{T}_2 \textbf{T}_1\boldsymbol{\gamma}.
\end{equation}
However~\cite{Balian1969}, since complex antisymmetric fermionic quadratic forms form a Lie algebra by Eq.~\eqref{int_rel}, there should exist complex antisymmetric matrices $\textbf{H}_1$ and $\textbf{H}_2$ such that $\textbf{T}_{1,2}=e^{\sigma \textbf{H}_{1,2}}$ and $e^{\sigma \textbf{H}}=e^{\sigma \textbf{H}_2}e^{\sigma \textbf{H}_1}$, provided $\mathcal{F}=\mathcal{F}_1\mathcal{F}_2$. Let us then consider the operator $\mathcal{F}$ and its associated matrix  
\begin{equation}
 \textbf{T}=e^{\sigma \bf{H}}\equiv\begin{pmatrix} \textbf{T}_{11} & \textbf{T}_{12}\\
    \textbf{T}_{21} & \textbf{T}_{22}
   \end{pmatrix}.
\end{equation}
According to the discussion above, we can find a factorization of $\mathcal{F}$ in the form
\begin{equation}
 \mathcal{F}=\mathcal{F}_1\mathcal{F}_2\mathcal{F}_3,
\end{equation}
such that $\mathcal{F}_3$ (resp. $\mathcal{F}_1$) only contains $\bmat{c}$ (resp. $\bmat{c}^{\dagger}$) operators. This implies the equation among matrices
\begin{equation}
 \textbf{T}=\text{exp}\left[{\sigma\begin{pmatrix} 0 & 0\\0 & \bf{X} \end{pmatrix}}\right]\text{exp}\left[{\sigma\begin{pmatrix} 0 & -\textbf{Y}^T\\\bf{Y} & 0 \end{pmatrix}}\right]\text{exp}\left[{\sigma\begin{pmatrix} \bf{Z} & 0\\0 & 0 \end{pmatrix}}\right]=\begin{pmatrix} e^{\bf{Y}}+\textbf{Z}e^{-\textbf{Y}^T}\bf{X} & \textbf{Z}e^{-\textbf{Y}^T} \\ e^{-\textbf{Y}^T}\bf{X} & e^{-\textbf{Y}^T}\end{pmatrix}
\end{equation}
which can be solved with the following result
\begin{equation}
\label{BB_dec}
 e^{-\textbf{Y}}=[\textbf{T}_{22}]^T,\quad \textbf{X}=\textbf{T}_{12}[\textbf{T}_{22}]^{-1},\quad \textbf{Z}=[\textbf{T}_{22}]^{-1}\textbf{T}_{21}.
\end{equation}
The antisymmetry of the matrix $\bf{Z}$ follows from the antisymmetry of $\textbf{T}_{21}\textbf{T}_{22}^T$, while the antisymmetry of $\bf{X}$ from the antisymmetry of $\textbf{T}_{12}\textbf{T}_{11}^T$. Furthermore we can also check that $\textbf{T}_{12}\textbf{T}_{21}^T+\textbf{T}_{11}\textbf{T}_{22}^T=\bf{1}$. The Eq.~\eqref{BB_dec} is the so-called Balian-Brezin factorization~\cite{Balian1969}. In the specific case discussed in Sec.~\ref{sec:1} one can then apply Eq.~\eqref{BB_dec} to the complex antisymmetric matrix
\begin{equation}
\textbf{H}=-it\begin{pmatrix} -\textbf{B} & -\textbf{A}\\
\textbf{A}& \textbf{B}\end{pmatrix}.
\end{equation}
%


\setcounter{equation}{0}
\renewcommand{\theequation}{A\arabic{equation}}

\end{document}